# Learning from Elders: Making an LLM-powered Chatbot for Retirement Communities more Accessible through User-centered Design

AUTHORS SECTION

**Li, Luna Xingyu** — Department of Biomedical Informatics and Medical Education at University of Washington, USA | lixy@uw.edu

**Chung, Ray-yuan** — Department of Biomedical Informatics and Medical Education at University of Washington, USA | raychung@uw.edu

**Chen, Feng** — Department of Biomedical Informatics and Medical Education at University of Washington, USA | fengc9@uw.edu

**Zeng, Wenyu** — Department of Biomedical Informatics and Medical Education at University of Washington, USA | wz34@uw.edu

**Jeon, Yein** — Department of Biomedical Informatics and Medical Education at University of Washington, USA | yeinj@uw.edu

**Zaslavsky, Oleg** — Digital Health Innovation Hub at the University of Washington, USA | ozasl@uw.edu

## ABSTRACT

Low technology and eHealth literacy among older adults in retirement communities hinder engagement with digital tools. To address this, we designed an LLM-powered chatbot prototype using a human-centered approach for a local retirement community. Through interviews and persona development, we prioritized accessibility and dual functionality: simplifying internal information retrieval and improving technology and eHealth literacy. A pilot trial with residents demonstrated high satisfaction and ease of use, but also identified areas for further improvement. Based on the feedback, we refined the chatbot using GPT-3.5 Turbo and Streamlit. The chatbot employs tailored prompt engineering to deliver concise responses. Accessible features like adjustable font size, interface theme and personalized follow-up responses were implemented. Future steps include enabling voice-to-text function and longitudinal intervention studies. Together, our results highlight the potential of LLM-driven chatbots to empower older adults through accessible, personalized interactions, bridging literacy gaps in retirement communities.

## KEYWORDS

Large language model; user-centered design; conversational AI; accessibility; retirement community

## INTRODUCTION

Retirement communities increasingly adopt digital tools to enhance residents' independence and quality of life (Wang et al., 2019). However, low technology and eHealth literacy among older adults may hinder their engagement. Our previous survey from a large retirement community in Seattle revealed that only 0–12% of residents actively use technologies available to them despite 85% expressing interest in learning more about such technologies. Barriers include complex interfaces, small font sizes, and lack of tailored guidance. Prior studies highlight chatbots as effective tools for improving health literacy and improving technology adoption (Meier et al., 2019; Mokmin & Ibrahim, 2021). Thus, we developed an LLM-powered chatbot that can be integrated into the retirement community's intranet platform to address these challenges through personalized, accessible interactions (Chung et al., 2024).

## PROJECT DESIGN AND SCOPE DEVELOPMENT

We adopted a human-centered design approach to develop the chatbot, ensuring the end-users' needs drive the design. At the outset, we conducted interviews with key stakeholders: a community board member from the retirement community and a current resident champion, in order to identify specific tasks that a chatbot could assist with. The interviews highlighted common struggles such as clicking through the app's menus to find information, checking dining hall menus, and adding events to a personal calendar. The resident stated a conversational agent could make these tasks more straightforward and believed current residents might be more inclined to explore new technologies with the addition of the conversational interface. The qualitative finding aligned with previous survey results. Together, these early user insights shaped the scope and priorities of our project.

We created 3 personas and 20 user stories to represent the diverse range of residents and their needs (details can be found in the GitHub repository). The personas and user stories encapsulated varying levels of tech experience, health conditions, and personal goals. For example, one persona represented a resident that is curious about learning technology but got frustrated by complexity, while another value staying informed about community events but has vision impairments requiring larger text. We also grounded our design using theoretical models such as the sociotechnical framework (Singh & Sittig, 2020), which recognized that successful implementation depends not just on the chatbot's technical design but also on social factors like training, support, and trust within the community. By acknowledging these factors, we planned for complementary activities like introductory workshops and ongoing user support.

In summary, our design and development process were iterative and user-centered: gathering input from older adults, building prototypes around their goals, and using established models to anticipate factors influencing acceptance.

## USER FEEDBACK AND PROTOTYPE UPDATE

Initially, to demonstrate the concept, we employed a customized version of OpenAI's ChatGPT as a prototype model (Brown et al., 2020). In the pilot evaluation session, two residents were invited to interact with the chatbot prototype following a think-aloud protocol. They were asked to complete two specific tasks, accessing the daily dining menu and adding a community event to their calendar. These tasks were among the most common use cases residents interacted with Miranet, as indicated in our initial interviews. These tasks were designed to (1) test the feasibility of using an AI-based conversational agent in a retirement community setting; and (2) gather user feedback for further chatbot refinement. The evaluation results were promising. Both residents successfully completed the assigned tasks with the use of the prototype, and survey responses indicated a high level of satisfaction with the ease of use and the usefulness of the chatbot. The residents appreciated the conversational style and the step-by-step prompts that guided them through tasks that, in the past, had been frustrating due to the complexity of the existing app. However, several key areas for improvement were identified. Participants suggested that the interface could be more accessible. For instance, larger font sizes and a high-contrast color scheme to accommodate visual impairments. They also expressed interest in a voice-to-text feature, which would allow them to interact with the chatbot without the need for extensive typing. Additionally, the need for clearer and more transparent privacy notices to address concerns about data usage was raised.

## INTERFACE IMPLEMENTATION

Following the pilot session, we implemented the full chatbot interface using OpenAI's GPT-3.5 Turbo as the language model backbone. The system was integrated via an API into a custom-built web interface specifically tailored for the target user group. We compared several LLMs performance including FLAN-T5, Phi3 and GPT-3.5 Turbo. GPT provided the best interaction. We employed guardrails via prompt engineering to ensure the model outputs are safe, concise and context-appropriate.

We used Streamlit to implement the interface for the chatbot. Figure 1 demonstrates the key features of the interface. This new interface was designed with a focus on accessibility and ease of use, incorporating the feedback obtained during the pilot phase. The interface features a clean, high-contrast layout with customizable font size and adjustable background color options, ensuring that residents with visual impairments can comfortably read and interact with the system. We enabled a simplified navigation. The home screen presents a set of predetermined topics—such as "Exercises for seniors," "Today's dining menu," and "How to setup phone reminders"—so that users can simply click on a topic of interest to start a conversation with the chatbot instead of having to type out complex queries. We also implemented a follow-up clickable prompts feature: once a topic is selected, the chatbot guides the interaction by providing three clickable follow-up questions generated by GPT-3.5. This design minimizes the need for extensive text input and makes the system approachable even for those with limited digital literacy.

Two primary functions guide the chatbot's operations. First, it assists with navigating community information. For example, residents can ask about today's lunch menu, upcoming events, or contact details for community staff. In these cases, the chatbot either retrieves information directly from the community internal database or provides easy-to-follow, step-by-step instructions for accessing the desired details. Moreover, the chatbot is designed to support users in learning about technology and health related topics. It can answer questions related to personal devices and applications—such as connecting to Wi-Fi on an iPad or scheduling a medication reminder on a smartphone—in simple steps. Moreover, it presents educational content on topics like online safety and telehealth services, always using clear, jargon-free language.

Safety and personalization features are integrated into the design. We engineered the chatbot to recognize potential emergency signals. For instance, if a user mentions severe symptoms or an urgent situation, the chatbot immediately advises them to seek professional help and displays emergency contact information. In addition, the chatbot is not designed to give medical advice. So, if a resident asks about personalized health questions such as "Can I eat an egg if I have hypertension?", the chatbot would reply with warnings and encourage users to consult a dietitian.

## NEXT STEP

Our next steps include further testing and refinement based on users' feedback, ensuring seamless integration with the existing community system and expanding features such as voice interaction. We aim to recruit 8 -12 users from the retirement community to evaluate the latest prototype. Additionally, we will compare the performance of multiple latest models including GPT-4o. Through these efforts, we aim to refine the user experience and foster greater adoption among older adults.

## CONCLUSION

The chatbot demonstrates potential in enhancing technology adoption and health literacy among older adults. Drawing from insights provided by older adults, we aim at refining accessibility features to improve user experience and foster broader adoption among this user group with specific needs. Continuous collaboration with community committees will also ensure sustained support and iterative improvements of the chatbot.

## GENERATIVE AI USE

We employed GPT-3.5 Turbo to generate conversational responses and follow-up questions. Outputs were validated for accuracy and simplicity by the research team. The authors assume full responsibility for the content.

## ACKNOWLEDGEMENTS

We thank the residents from the retirement community, Dr. Oleg Zaslavsky, and Prof Annie T Chen for their support.

## CODE ACCESSIBILITY

https://github.com/chenfeng1234567/llm_chatbot

## SUPPLEMENTARY

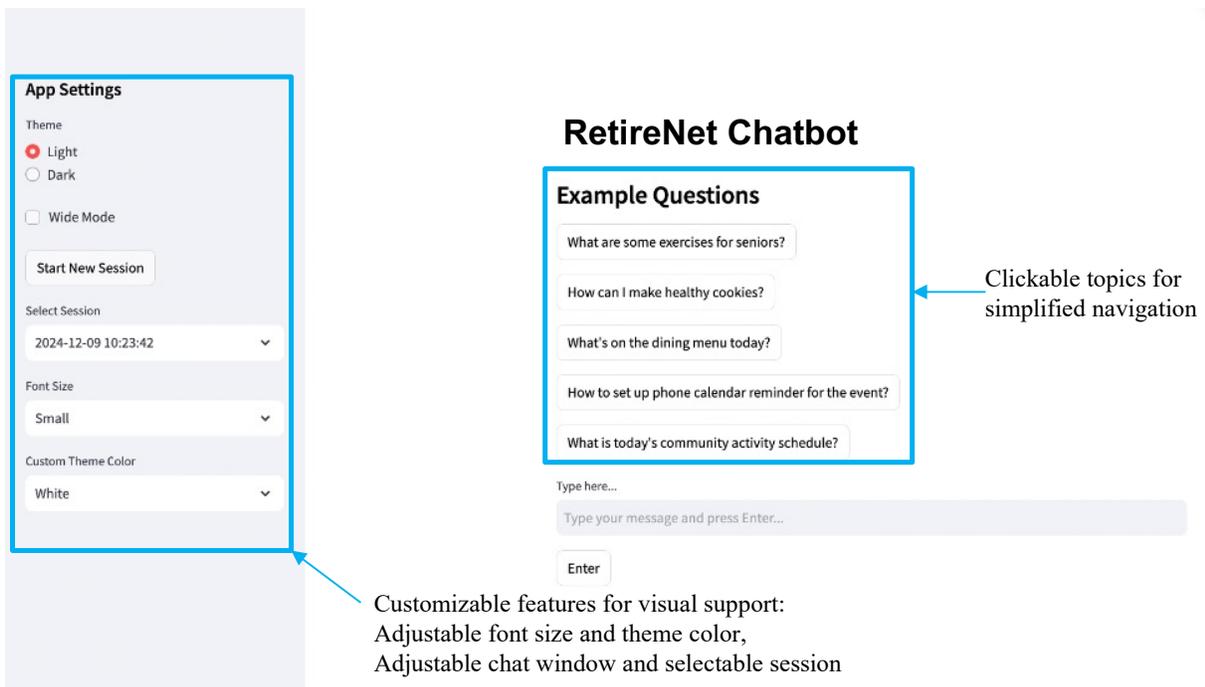

Figure 1. Screenshot of the implemented chatbot interface. The accessible features are highlighted in blue boxes.